\documentclass{llncs}

\usepackage{enumitem}
\usepackage{color}
\usepackage{cite}
\usepackage{graphicx}
\usepackage{algorithm2e}
\usepackage{adjustbox}

\begin{document}

\title{Exploration of Pattern-Matching Techniques for Lossy Compression on Cosmology Simulation Data Sets}

\author{Dingwen Tao\inst{1} \and Sheng Di\inst{2} \and Zizhong Chen\inst{1} \and Franck Cappello\inst{2,3}}

\institute{
University of California, Riverside, CA, USA \\
\{dtao001, chen\}@cs.ucr.edu
\and Argonne National Laboratory, IL, USA \\
\{sdi1, cappello\}@anl.gov
\and University of Illinois at Urbana-Champaign, IL, USA
}

\maketitle

\begin{abstract}
Because of the vast volume of data being produced by today's scientific simulations, lossy compression allowing user-controlled information loss can significantly reduce the data size and the I/O burden. However, for large-scale cosmology simulation, such as the Hardware/Hybrid Accelerated Cosmology Code (HACC), where memory overhead constraints restrict compression to only one snapshot at a time, the lossy compression ratio is extremely limited because of the fairly low spatial coherence and high irregularity of the data. In this work, we propose a pattern-matching (similarity searching) technique to optimize the prediction accuracy and compression ratio of SZ lossy compressor on the HACC data sets. We evaluate our proposed method with different configurations and compare it with state-of-the-art lossy compressors. Experiments show that our proposed optimization approach can improve the prediction accuracy and reduce the compressed size of quantization codes compared with SZ. We present several lessons useful for future research involving pattern-matching techniques for lossy compression.
\end{abstract}
\section{Introduction}

Because of ever-increasing parallel execution scale, today's scientific simulations are producing volumes of data too  large to be accommodated in storage systems. The limitation comes from the limited storage capacity and I/O bandwidth of parallel file systems in production facilities. Cosmology simulations such as the Hardware/Hybrid Accelerated Cosmology Code (HACC) \cite{habib2016hacc} are typical examples of parallel executions facing this issue. HACC solves an N-body problem involving domain decomposition, a medium-/long-range force solver based on a particle-mesh method, and a short-range force solver based on a particle-particle/particle-mesh algorithm. According to cosmology researchers, the number of particles to simulate can be up to 3.5 trillion in today's simulations (and even more in the future), which leads to 60 PB of data to store; yet a system such as the Mira  supercomputer has only 26 PB of file system storage. Currently, HACC users rely on decimation in time, storing only a fraction of the simulation snapshots, to reduce the pressure on the storage system. A reduction factor of 80\% to 90\% is commonly used. At exascale, temporal decimation will not be enough to address the limitations of the storage system: snapshots will be so large (each in the range of 5 PB) that the time to store each snapshot (83 minutes on a storage system offering a sustained bandwidth of 1 TB/s) will become a serious problem. HACC is not a special case. As indicated by \cite{glecler}, nearly 2.5 PB of data were produced by the Community Earth System Model for the Coupled Model Intercomparison Project (CMIP) 5, which further introduced 170 TB of postprocessing data submitted to the Earth System Grid \cite{esg}. Estimates of the raw data requirements for the CMIP6 project exceed 10 PB \cite{baker}. At exascale, storing each full snapshot in this case would also take too long, however, so that on-line/in situ compression of each snapshot is needed. 

In this paper, we explore pattern-matching techniques for lossy compression, focusing  on individual snapshots of the scientific data sets produced by cosmology simulations. 
Because of the constraints of memory consumption, we cannot leverage the smoothness of a particle's trajectory (such as smoothness along the time dimension) to reduce the data size; hence, we must  perform compression on individual snapshots. Unlike the mesh data produced by conventional simulations, such as fluid dynamics, the data of particles in cosmology simulations, such as coordinate and velocity data, are stored in separate 1D arrays. In the HACC application, the indices of each 1D array are kept consistent for the same cosmology particle. Specifically, the HACC simulation data contains six 1D arrays:  three coordinate fields (xx, yy, zz) and three velocity fields (vx, vy, vz). Because of the lack of correlation between adjacent particles in the HACC data set, state-of-the-art lossy compressors, such as FPZIP [16], ZFP [15] and SZ [9, 20], reach relatively low compression ratios/factors (2 to 5 with the error bound set to $10^{-4}$).

The rest of the paper is organized as follows.
In Section \ref{sec: problem}, we formulate the data compression problem based on cosmology simulation data sets and the assessment of several state-of-the-art lossy compressors on the HACC data sets. In Section \ref{sec: pattern}, we discuss the well-known dictionary-based lossless compression algorithm LZ77 and propose our pattern-matching-based optimization method for SZ lossy compression for low spatial coherence and highly irregular data, such as the velocity variables in the HACC data sets. In Section \ref{sec: evaluation}, we evaluate the compression ratios of our proposed optimization method and compare it with one variant of the SZ lossy compressor. We discuss  related work in Section \ref{sec: related} and provide  conclusions in Section \ref{sec: conclusion}.

\section{Related Work}
\label{sec: related}
Data compression has been extensively studied for decades and can be split into two categories: lossless compression and lossy compression. The main limiation of the lossless compressors (such as GZIP \cite{gzip}) is their fairly low compression ratio on scientific data sets composed of floating-point values, as confirmed by \cite{ratana, sz16,sz17}. 

\begin{figure}[t]
\centering
\includegraphics[scale=0.56]{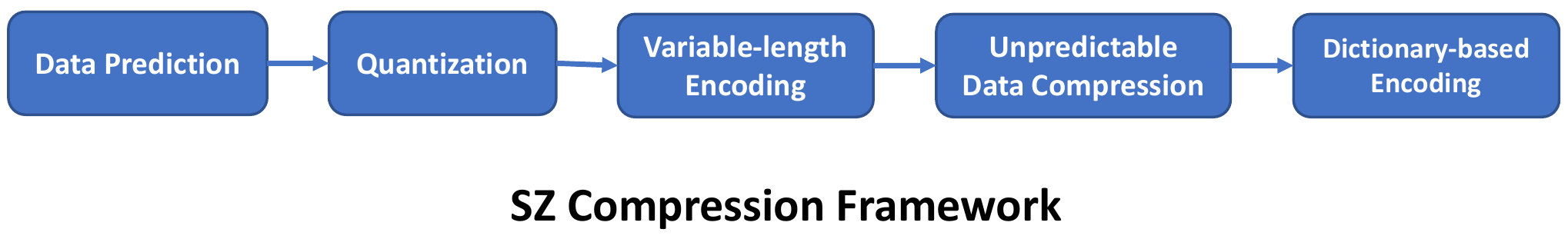}
\caption{Overview of SZ lossy compression algorithm.}
\label{fig:sz}
\vspace{-4mm}
\end{figure}

Recently, many lossy compressors have been designed and implemented for scientific data. Most of them are designed for mesh data sets, which are expected to have strong coherence among the nearby data in the data set,  but the quality of their compression declines on cosmology simulation data sets. For example, SZ \cite{sz16, sz17} has five main steps including (1) data prediction for each point by its preceding neighbors in the multidimensional space, (2) error-controlled linear quantization, (3) customized Huffman coding  \cite{huffman} (i.e., variable-length encoding) to shrink the data size significantly, (4) unpredictable data compression, and (5) customized LZ77 coding (i.e., dictionary-based encoding). The compression framework of SZ is shown in Figure \ref{fig:sz}. ZFP \cite{zfp} splits the whole data set into many small blocks with an edge size of 4 along each dimension and compresses the data in each block separately by a series of carefully designed steps (including alignment of exponent, orthogonal transform, fixed-point integer conversion, and binary representation analysis with bit-plane encoding). FPZIP \cite{fpzip} adopts predictive coding and  ignores insignificant bit planes in the mantissa based on the analysis of IEEE 754 binary representation \cite{ieee2008754}. SSEM \cite{ssem} splits data into a high-frequency part and low-frequency part by wavelet transform \cite{wavelet} and then uses vector quantization and GZIP. ISABELA \cite{isabela} sorts the data and then performs the data compression by B-spline interpolation; but it has to store an extra index array to record the original location for each point, and it suffers significantly from low compression ratio. Compression of particle simulation data sets has also been studied for years, but most of the methods proposed are based on smooth temporal trajectory of the same particles, which requires loading/keeping multiple snapshots during the compression/simulation \cite{Yang-sc1999, ed2006, Kumar2013, dct, numarck}. Thus, they are not suitable for extremely large-scale simulation in which only one snapshot is allowed to be loaded into the memory. Omeltchenko et al. \cite{cpc2000} proposed a lossy compression method (called CPC2000 in this paper) that does not rely on temporal coherence and relies on only a single snapshot. Its main steps involve reorganizing all particles in the space onto a zigzag-similar space-filling curve \cite {zigzag}, sorting the particles based on the R-indices by a radix-similar sorting method in each block, and compressing the difference of the adjacent indices by adaptive variable-length coding.
\section{Problem Formulation}
\label {sec: problem}

Scientific data compression algorithms can be classified into two categories: lossless compression and lossy compression. The main limitation of lossless compressors is their limited data reduction capability, that is, up to 2:1 in general \cite{ratana} and even lower on cosmology simulation simulation data sets. In this work, therefore, we focus on lossy compression methods for cosmology simulations.

Cosmology simulations generate multiple snapshots. Because of considerations of memory consumption, we focus on single-snapshot compression without using temporal coherence in this work. Such simulations contain many variables each representing one data field of particles. In the HACC simulation data considered in this study, the variables are stored in separate 1D arrays. Specifically, each snapshot of HACC simulation contains six single-precision floating-point variables:  $xx$, $yy$, $zz$, $vx$, $vy$, and $vz$. The first three indicate coordinate information, and the other three indicate velocity along the three dimensions. The six variables are stored in separate  floating-point arrays. Unlike regular multidimensional mesh data, the particle elements in each array are allowed to be reordered in the reconstructed data set, whereas the locations or indices of the elements with regard to the same particle must be consistent across arrays.

The main objective of our work is to optimize the single-snapshot lossy compression ratio for cosmology simulation data sets, provided that the compression errors are controlled within a user-specified bound for each data point. \textit{Compression ratio} is the ratio of the original data size to the compressed data size. Table \ref{tab: compratio} shows the compression ratios of several state-of-the-art lossy compressors on the HACC data sets under the value-range-based relative error bound $10^{-4}$, denoted by $eb_{rel} = 10^{-4}$. The version of the SZ lossy compressor we focus on in this work is ``SZ-LV'', which is based on the last-value prediction model. Note that for CPC2000, ZFP, and SZ, we use the absolute error bounds computed based on $eb_{rel} = 10^{-4}$ and the value range of each variable; for FPZIP, we set the number of retained bits to 21 as approximate $eb_{rel} = 10^{-4}$ for all the variables. The SZ lossy compressor has higher compression ratios on the coordinate variables (i.e., $xx, yy, zz$) than on the velocity variables (i.e., $vx, vy, vz$). Therefore, in this work we focus on optimizing the prediction accuracy and compression ratios based on SZ lossy compression for the velocity variables in the HACC data.

\begin{table}[]
\centering
\caption{Compression ratios of different variables with different compressors on HACC data sets under value-range-based relative error bound $10^{-4}$.}
\label{tab: compratio}
\begin{tabular}{|l|c|c|c|c|c|c|}
\hline
\textbf{Compressor} & \textbf{xx} & \textbf{yy} & \textbf{zz} & \textbf{vx} & \textbf{vy} & \textbf{vz} \\ \hline
CPC2000             & 7.1         & 7.1         & 7.1         & 2.3    &   2.3  & 2.3 \\ \hline
FPZIP               & 5.8         & 5.7         & 4.4         & 2.2    &   2.2  & 2.2 \\ \hline
ZFP                 & 2.3         & 2.3         & 2.2         & 2.3   &  2.3    & 2.3 \\ \hline
SZ                  & 8.2         & 8.3         & 5.9         & 4.0   &  4.0    & 4.0 \\ \hline
\end{tabular}
\end{table}

\newpage
\section{Pattern-Matching Techniques for Lossy Compression}
\label{sec: pattern}

In this section, we first discuss the well-known dictionary-based lossless compression algorithm Lempel-Ziv 77 (LZ77). It can encode a sequence of symbols and compress the input source by using the information of recently frequent consecutive symbols. Inspired by LZ77's classic idea, we then propose our pattern-matching-based lossy compression method, called SZ-PM. Because of different input sources, we propose many tailored designs for dealing with lossy compression and floating-point scientific data.

\subsection{LZ77: string matching based lossless compression}

\begin{algorithm}[H]
 \While{look-ahead buffer is not empty}{
  go backwards in search buffer to find longest match of the look-ahead buffer\;
  \eIf{match found}{
   output (offset, length, next symbol in look-ahead buffer)\;
   shift sliding window by length+1\;
   }{
   output (0, 0, first symbol in look-ahead buffer)\;
   shift sliding window by 1\;
  }
 }
 \label{algo:lz77}
 \caption{Pseudo code of the LZ77 algorithm}
\end{algorithm}

\begin{figure}[t]
\centering
\includegraphics[scale=0.44]{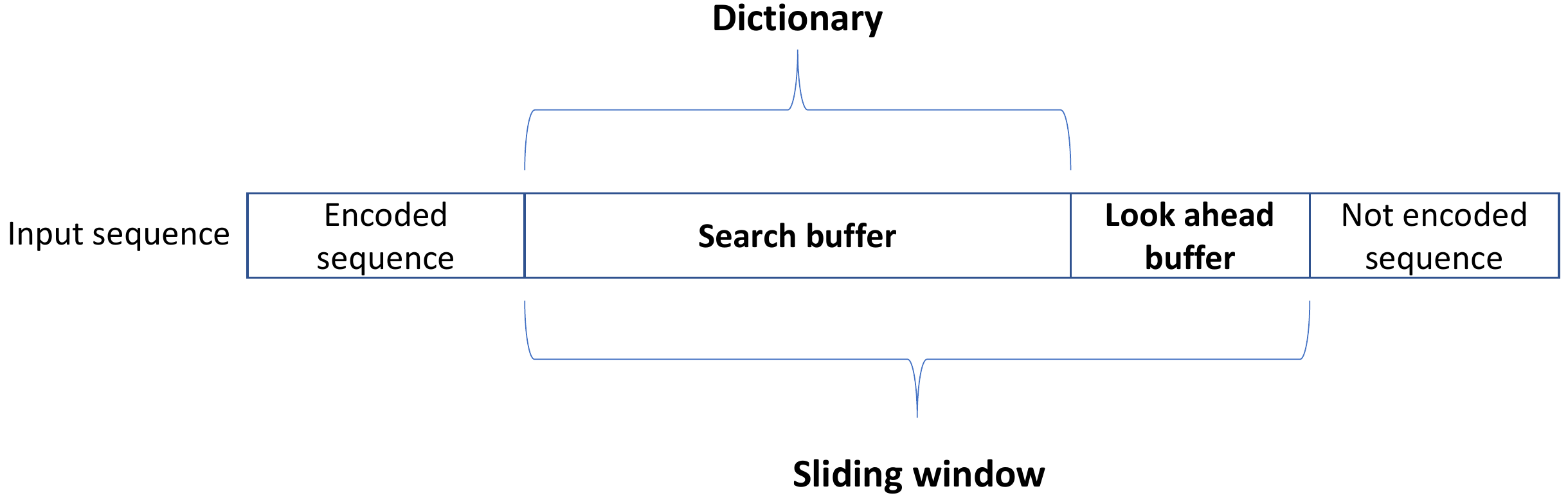}
\caption{Overview of LZ77 lossless compression algorithm. }
\label{fig:lz77}
\vspace{-4mm}
\end{figure}

The Lempel-Ziv 77 (LZ77) lossless compression algorithm is the first Lempel-Ziv compression algorithm. Unlike scientific data compression, LZ77  is designed for encoding a sequence of symbols byte by byte based on a dictionary  constructed from a portion of the recently encoded sequence. Specifically, LZ77 encodes the input sequence through a sliding window  composed of two buffers, a search buffer and a look-ahead buffer, as shown in Figure \ref{fig:lz77}. The search buffer contains the most recently compressed symbols, while the look-ahead buffer contains multiple uncompressed symbols. The algorithm  searches the longest prefix of the look-ahead buffer that is also contained in the search buffer. The details of LZ77  are shown in Algorithm \ref{algo:lz77}. The LZ77 algorithm searches all the consecutive symbols in the search buffer to identify whether these symbols match the consecutive symbols in the look-ahead buffer. The offset in the algorithm represents the distance of the longest match's first symbol (in the search buffer) from the look-ahead buffer, and length represents the length of the longest match. Therefore, the general idea of LZ77  is to save  storage by using the information from the recent  symbol sequences based on a string-matching approach. It inspires us to design a similar matching technique for lossy scientific data compression.

\subsection{SZ-PM: pattern-matching-based lossy compression} 
We  propose a pattern-matching-based lossy compression method called SZ-PM. The idea of pattern matching is similar to the string matching idea used in LZ77. It is also designed to use the information of recent floating-point sequences with similar pattern in order to improve the prediction accuracy and compression ratio of SZ lossy compression for irregular data. Unlike the lossless compression algorithm for symbols (one byte per symbol), however, the lossy compression for scientific data is designed mainly for single/double floating-point data (4/8 bytes per value) and can tolerate compression errors within user-controlled error bounds. Therefore, we can design many tailored features for the pattern-matching method.

Let us first define necessary notations. Similar to LZ77, our algorithm also maintains two buffers in the sliding window during the compression: a search buffer and a look-ahead buffer. Let the search buffer size  be $m$ and the look-ahead buffer size  be $n$. Here the buffer size represents the number of data points in the buffer.  Let the $m$ compressed data points in the search buffer  be $\{s_1, s_2, ..., s_m\}$ and the $n$ uncompressed data points in the look-ahead buffer  be $\{l_1, l_2, ..., l_n\}$. Let the $m-n+1$ sequences with length of $n$ in the search buffer to be $X_1, X_2, ..., X_{m-n+1}$, where $X_1 = \{s_1, ..., s_n\}, X_2 = \{s_2, ..., s_{n+1}\}, ..., X_{m-n+1} = \{s_{m-n+1}, ..., s_m\}$. Let the one sequence with length of $n$ in the search buffer  be $Y = \{l_1, l_2, ..., l_n\}$.

We now describe our tailored designs of pattern matching for lossy compression and scientific data.
For compression, 
(1) we fix the length of matching sequences to be the size of look-ahead buffer (i.e., $n$). In other words, we attempt to identify the most similar sequence in the search buffer for the whole look-ahead buffer with length  $n$.
(2) We sort the $n$ data points in each sequence, including $X_1$, $X_2$, ..., $X_{m-n+1}$ from the search buffer and $Y$ from the look-ahead buffer. 
(3) For each sorted sequence, we subtract the mean value of the sequence from each value. In  other words, we shift the sequence by its mean value as $X = (x_1 - \overline{X}, x_2 - \overline{X}, ..., x_n - \overline{X})$ and $Y = (y_1 - \overline{Y}, y_2 - \overline{Y}, ..., y_n - \overline{Y})$, where $\overline{X} = \frac{1}{n} \sum\limits_{i=1}^n x_i$ and $\overline{Y} = \frac{1}{n} \sum\limits_{i=1}^n y_i$.
(4) We attempt to match the sequences from the search buffer for the look-ahead buffer, but we relax the ``matching'' condition. Specifically, the matching condition of LZ77 algorithm is that two symbol sequences are exactly the same; but in our algorithm we define two shifted floating-data sequences $X = (x_1, x_2, ..., x_n)$ and $Y = (y_1, y_2, ..., y_n)$ as ``matched'' if $(\sum\limits_{i=1}^n |x_i - y_i|^p)^{1/p} < \theta$, where $\theta$ is a given threshold, $X$ is one shifted sequence from the search buffer, and $Y$ is the shifted sequence of the look-ahead buffer. Note that the search buffer can have multiple matched sequences.
(5) We pick the matched sequence $X^*$ with the smallest distance from the multiple matched sequences as the most similar sequence for $Y$. We denote the values in $X^*$ by $\{x^*_1, x^*_2, ..., x^*_n\}$. We name this matching process as ``pattern matching'' and the sequence $X^*$ as the ``pattern matched sequence'' for $Y$. 
(6) We always shift the sliding window by length of $n$ after we go over the $m-n+1$ sequences in the search buffer. Unlike LZ77, we also shift the sliding window by length of $n$, even if we cannot find a matched sequence under the threshold $\theta$. 
(7) We use $X^*$ as the prediction sequence for $Y$, if the pattern matched sequence can be found. Specifically, we take $x^*_i - \overline{X^*}$ as the prediction value for $y_i - \overline{Y}$ of data point $i$. We use SZ's original prediction model proposed in \cite{sz17} to generate the prediction values for $Y$, if no matched sequence exists in the search buffer. Therefore, we must use an extra bit, denoted by $bit_{predmd}$, to represent the prediction method of each sequence. For example, we use $bit_{predmd} = 0$ to indicate that the sequence is predicted by pattern-matching method and $bit_{predmd} = 1$ to indicate that the sequence is predicted by SZ's original prediction model. 
(8) Similar to LZ77, if the sequence is predicted by the pattern-matching method, we still have to store the offset; but we do not need to store the length due to the fixed length. We also have to store the mean value of $Y$ in order to reconstruct the data during the decompression.
(9) We use the linear quantization method and the customized Huffman coding proposed in \cite{sz17} to encode the differences between prediction values and real values for $Y$ and compress the quantization codes based on the user-set error bound. Because of space limitations, we do not describe them in detail here.

For decompression, we use the same decompression method proposed in \cite{sz17} to construct the differences between prediction values and real values for each sequence. For example, in decompressing the sequence $Y$, we denote the difference of data point $i$ in $Y$ by $y_i^{diff}$.  We then construct the prediction values of $Y$ by its corresponding prediction method known from $bit_{predmd}$. If $bit_{predmd}$ indicates $Y$ is predicted by SZ's original prediction model during the compression, we construct its prediction values using the same process described in \cite{sz17}; if $bit_{predmd}$ indicates $Y$ is predicted by the pattern-matching approach during the compression, we  use the stored offset and mean value to construct the prediction values. Specifically, we can construct the prediction value of data point $i$ by $y_i^{pred} = x^*_i - \overline{X^*} + \overline{Y}$, where $X^*$ is the pattern-matched sequence that has already been decompressed. After constructing the prediction values for $Y$, we can reconstruct the value of data point $i$ by $y_i^{decomp} = y_i^{pred} + y_i^{diff}$.

\begin{figure}[t]
\centering
\includegraphics[scale=0.5]{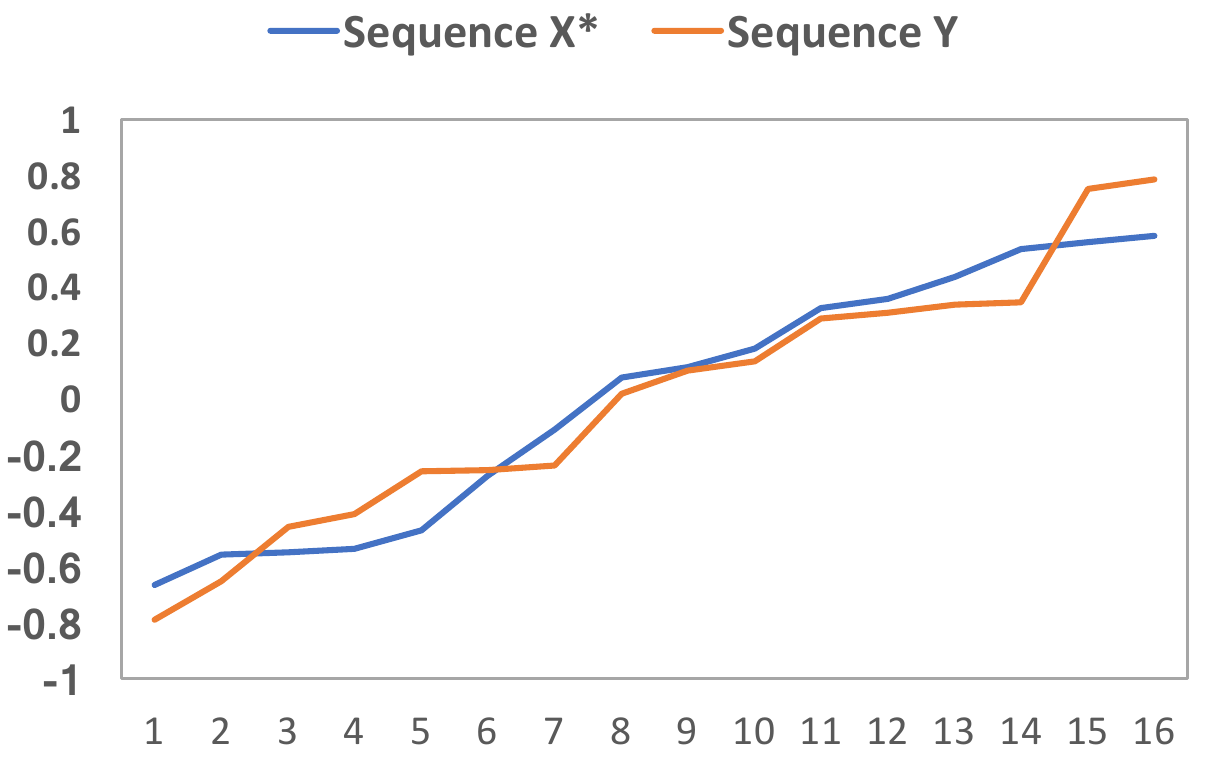}
\vspace{-4mm}
\caption{Example of two pattern matched sequences after sorting and shifting.}
\label{fig:matched}
\end{figure}

Algorithm \ref{algo:pattern} shows the pseudo code of our proposed pattern-matching-based lossy compression method. Figure \ref{fig:matched} shows an example of two pattern-matched sequences transformed by sorting and shifting. We have several remarks here. (1) For our matching condition, we treat the two $n$-length floating-point sequences as two data points in the $n$-dimensional space and define them as ``matched'' if their distance in Lp norm is smaller than the threshold $\theta$. According to \cite{chan1999efficient}, we set $\theta$ to $0.5$ of the search buffer size. (2) From our initial study we find that $p>1$ cannot reduce the size of the compressed quantization codes on the HACC data; hence we set $p=1/2$ in our algorithm and the following evaluation. (We will research the optimal $p$ in the future.) (3) As a result of the sorting process, the reconstructed data is recorded in one sequence. But as described in Section \ref{sec: problem}, the particle elements in each 1D array are allowed to be reordered in the reconstructed data sets. Hence, we do not have to extra storage to record the initial index information. (4) We use extra memory space to sort and shift the sequences without any modifications of the original data. The reason for sorting and shifting is to increase the possibility of matching sequences due to the high irregularity of the data and the relatively large value range of the floating-point data.

\begin{algorithm}[H]
 \While{look-ahead buffer is not empty}{
  sequence $Y$ is composed of the $n$ data points of the look-ahead buffer\;
  search buffer contains $m-n+1$ sequences  $\{X_1, X_2, ..., X_{m-n+1}\}$\;
  sort each sequence including $X_1, X_2, ..., X_{m-n+1}$ and  $Y$\;
  compare sorted $Y$ with $\{X_1, X_2, ..., X_{m-n+1}\}$ and find sequence $X^*$ with the smallest distance (in Lp norm) from Y, i.e., $dist(X^*, Y)$\;
  \eIf{$dist(X^*, Y) < \theta$}{
  $bit_{predmd}$  = 0\;
  store (offset, mean value $\overline{Y}$)\;
  prediction values of $Y$ are calculated by
  $y_i^{pred} = x^*_i - \overline{X^*} + \overline{Y}$\;
  }{
  $bit_{predmd}$  = 1\;
  use SZ's original prediction model to predict values of $Y$\;
  }
  calculate differences between real value $y_i$ and prediction value $y_i^{pred}$\;
  encode differences using linear quantization method based on user-set error bound\;
  compute and record decompressed value\;
  shift sliding window by length of $n$\;
 }
 compress linear quantization codes using Huffman coding\;
 compress unpredictable data by SZ's binary representation analysis\;
 
 \caption{Pseudo code of SZ-PM algorithm}
 \label{algo:pattern}
\end{algorithm}

\section{Empirical Evaluation}
\label{sec: evaluation}

In this section, we evaluate our proposed lossy compression method, SZ-PM, on the velocity variables in the HACC data sets, and we compare it with the SZ lossy compressor \cite{sz17}. Note that the SZ lossy compressor we evaluate in this study is a variant of the original SZ. It first splits the original data into multiple segments. The segment size is consistent with the look-ahead buffer size. It then performs a sorting within each segment. After that, it conducts the original SZ compression on the transformed data. The reason of using this variant version is that we want to evaluate the effects of the pattern-matching method  without impact from the sorting technique and to compare SZ and SZ-PM in a fair level.

\begin{figure}[t]
\centering
\includegraphics[scale=0.46]{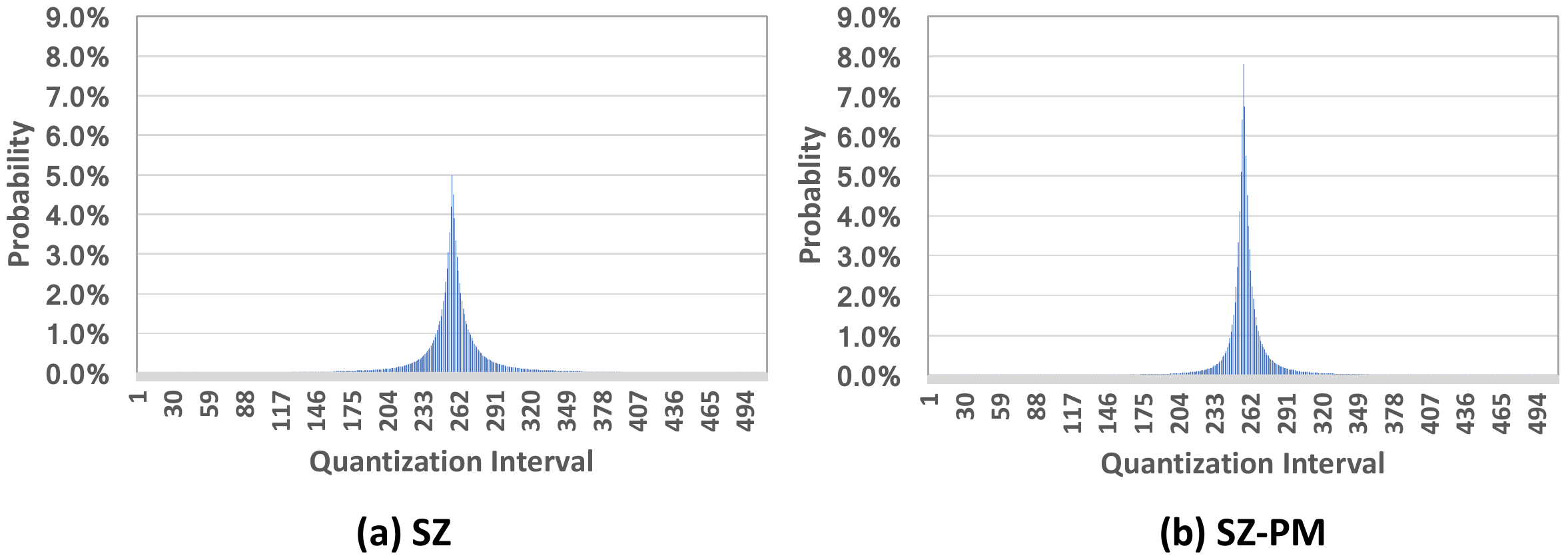}
\vspace{-2mm}
\caption{Distribution produced by linear quantization encoder in (a) SZ and (b) SZ-PM on the velocity variable $vx$ in the HACC data sets with 511 quantization intervals.}
\label{fig:distribution}
\vspace{-4mm}
\end{figure}

As described in \cite{sz17}, the distribution produced by linear quantization encoder can significantly affect the performance of Huffman coding \cite{huffman}. Generally speaking, the more concentrated the distribution, the higher the compression ratio that the Huffman coding can achieve. Figure \ref{fig:distribution} shows the distributions produced by linear quantization encoder in the SZ and our proposed SZ-PM lossy compression method on the velocity variable $vx$ in the HACC data sets. Note that we use a $10^{-4}$ value-range-based relative error bound and 511 quantization intervals. Based on our observation, 511 quantization intervals can cover more than 99.9\% data points during the linear quantization in this case. The figure illustrates that our proposed SZ-PM can improve the prediction accuracy and make the distribution of quantization code more concentrated. (We will show the incremental results in detail later.)

\begin{table}[]
\vspace{-6mm}
\centering
\caption{Evaluation of our proposed SZ-PM on the velocity variable $vx$ in the HACC data sets with different sizes of sorting/matching sequence.}
\label{tab:results}
\begin{adjustbox}{width=\columnwidth}
\begin{tabular}{|l|c|c|c|c|c|c|c|}
\hline
                   & \textbf{\begin{tabular}[c]{@{}c@{}}Size of\\ Quantization Code\\ (bits/value)\end{tabular}} & \textbf{\begin{tabular}[c]{@{}c@{}}Size of\\ $bit_{predmd}$\\ (bits/value)\end{tabular}} & \textbf{\begin{tabular}[c]{@{}c@{}}Ratio of\\ PM Sequence\\ (\%)\end{tabular}} & \textbf{\begin{tabular}[c]{@{}c@{}}Size of\\ Offset\\ (bits/value)\end{tabular}} & \textbf{\begin{tabular}[c]{@{}c@{}}Size of\\ Mean Value\\ (bits/value)\end{tabular}} & \textbf{\begin{tabular}[c]{@{}c@{}}Overall\\ Bit-rate\\ (bits/value)\end{tabular}} & \textbf{\begin{tabular}[c]{@{}c@{}}Compression\\ Ratio\end{tabular}} \\ \hline
\textbf{CPC2000}   & /                                                                                           & /                                                                                        & /                                                                              & /                                                                                & /                                                                                    & 13.9                                                                               & 2.30                                                                 \\ \hline
\textbf{SZ(8)}     & 7.31                                                                                        & /                                                                                        & /                                                                              & /                                                                                & /                                                                                    & 7.3                                                                                & 4.38                                                                 \\ \hline
\textbf{SZ-PM(8)}  & 5.45                                                                                        & 1/8                                                                                      & 99.6\%                                                                         & 1.25                                                                             & 3.98                                                                                 & 10.8                                                                               & 2.96                                                                 \\ \hline
\textbf{SZ(16)}    & 6.75                                                                                        & /                                                                                        & /                                                                              & /                                                                                & /                                                                                    & 6.8                                                                                & 4.74                                                                 \\ \hline
\textbf{SZ-PM(16)} & 6.01                                                                                        & 1/16                                                                                     & 93.1\%                                                                         & 0.58                                                                             & 1.86                                                                                 & 8.5                                                                                & 3.76                                                                 \\ \hline
\textbf{SZ(32)}    & 6.16                                                                                        & /                                                                                        & /                                                                              & /                                                                                & /                                                                                    & 6.2                                                                                & 5.19                                                                 \\ \hline
\textbf{SZ-PM(32)} & 6.07                                                                                        & 1/32                                                                                     & 66.3\%                                                                         & 0.04                                                                             & 0.66                                                                                 & 6.8                                                                                & 4.71                                                                 \\ \hline
\end{tabular}
\end{adjustbox}
\vspace{-7mm}
\end{table}

Table \ref{tab:results} shows the experimental results of our evaluation for SZ-PM on the HACC data sets. In the experiments, we set the search buffer size to 1024; hence, we need to use 10 bits ($2^{10} = 1024$) to represent the offset value for each sequence that is predicted by the pattern-matching method during compression. We test SZ-PM with different configurations of three look-ahead buffer sizes:  8, 16, and 32. The size of each category presented in the table is the atomized size (i.e., bits per value). Note that the original data type of the HACC data is single floating-point (i.e., 32 bits per value); hence, the compression ratio can be calculated by $32/overall\_size$. The number in each bracket represents the segment size/sequence size; for example, SZ(8) means that the segment size used for sorting in SZ is 8, and SZ-PM(8) means that the length of sequence used in the pattern matching is 8. The column ``Ratio of PM Sequence'' means the ratio of the sequences predicted by the pattern matching during compression. 

We make several observattions from Table \ref{tab:results}. 
(1) SZ-PM can improve the prediction accuracy and reduce the size of the compressed quantization codes.
(2) The shorter the matching sequence is, the more accurately the SZ-PM can predict.
(3) For SZ-PM, the shorter the matching sequence is, the smaller the compressed quantization codes will be; 
however, for SZ, on the contrary, the longer the segment is, the smaller the compressed quantization codes will be.
(4) The longer the matching sequence is, the less the storage overhead that the offset and mean values will have.
(5) The reduced size of the compressed quantization codes, achieved from the improvement of the prediction accuracy by SZ-PM, is counteracted by the incremental overhead of storing offset and mean values.

From these observations, we derive some useful lessons for future research with respect to the pattern-matching techniques in lossy compression as follows. (1) Our proposed pattern-matching technique can enhance the prediction accuracy and reduce the size of compressed quantization codes, but the improvement is not enough to cover the extra overhead introduced by storing offset and mean values. (2) We should further improve the prediction accuracy using a more advanced pattern-matching technique. (3) We should reduce/eliminate the extra overhead of offset and mean values, especially the mean values of floating-point data type. For example, we may shift the sequence by the value of the first element in the sequence; consequently, we do not need to store the mean values. (4) Currently, we consider reordering only one variable in the HACC data sets. In  future research, we need to consider the impact of reordering one variable to the other variables, since we have to make all the variables consistent.
\section{Conclusion}
\label{sec: conclusion}
\vspace{-2mm}
In this work, we explored pattern-matching techniques for lossy compression based on the SZ compressor. The experiments demonstrate that our proposed optimization method, SZ-PM, can improve the prediction accuracy and reduce the size of compressed quantization codes on the HACC velocity data, but the compression ratio cannot be improved because of storing extra information. We plan to explore ways to improve the prediction accuracy with the pattern-matching technique and to reduce the storage of extra information.
\section*{Acknowledgments}
\scriptsize
This material is based upon work supported by the U.S. Department of Energy, Office of Science, under contract number DE-AC02-06CH11357.
The submitted manuscript has been created by UChicago Argonne, LLC, Operator of Argonne National Laboratory (``Argonne''). Argonne, a U.S. Department of Energy Office of Science laboratory, is operated under Contract No. DE-AC02-06CH11357. The U.S. Government retains for itself, and others acting on its behalf, a paid-up nonexclusive, irrevocable worldwide license in said article to reproduce, prepare derivative works, distribute copies to the public, and perform publicly and display publicly, by or on behalf of the Government.  The Department of Energy will provide public access to these results of federally sponsored research in accordance with the DOE Public Access Plan. http://energy.gov/downloads/doe-public-access-plan.

\bibliographystyle{splncs03}
\bibliography{bib/refs} 

\begin{thebibliography}{10}
\providecommand{\url}[1]{\texttt{#1}}
\providecommand{\urlprefix}{URL }

\bibitem{dct}
Ahmed, N., Natarajan, T., Rao, K.R.: Discrete cosine transform. IEEE
  Transactions on Computers  100(1),  90--93 (1974)

\bibitem{baker}
Baker, A.H., Xu, H., Dennis, J.M., Levy, M.N., Nychka, D., Mickelson, S.A.,
  Edwards, J., Vertenstein, M., Wegener, A.: A methodology for evaluating the
  impact of data compression on climate simulation data. In: HPDC'14. pp.
  203--214 (2014)

\bibitem{esg}
Bernholdt, D., Bharathi, S., Brown, D., Chanchio, K., Chen, M., Chervenak, A.,
  Cinquini, L., Drach, B., Foster, I., Fox, P., et~al.: {The Earth System
  Grid}: Supporting the next generation of climate modeling research.
  Proceedings of the IEEE  93(3),  485--495 (2005)

\bibitem{chan1999efficient}
Chan, K.P., Fu, A.W.C.: Efficient time series matching by wavelets. In:
  Proceedings of the 15th International Conference on Data Engineering. pp.
  126--133. IEEE (1999)

\bibitem{zigzag}
Chanussot, J., Lambert, P.: Total ordering based on space filling curves for
  multivalued morphology. Computational Imaging and Vision  12,  51--58 (1998)

\bibitem{numarck}
Chen, Z., Son, S.W., Hendrix, W., Agrawal, A., Liao, W., Choudhary, A.N.:
  {NUMARCK:} machine learning algorithm for resiliency and checkpointing. In:
  {SC} 2014. pp. 733--744 (2014)

\bibitem{ieee2008754}
Committee, I.S., et~al.: 754-2008 {IEEE} standard for floating-point
  arithmetic. IEEE Computer Society Std  2008 (2008)

\bibitem{wavelet}
Daubechies, I.: The wavelet transform, time-frequency localization and signal
  analysis. IEEE transactions on information theory  36(5),  961--1005 (1990)

\bibitem{gzip}
Deutsch, L.P.: {GZIP} file format specification version 4.3  (1996)

\bibitem{sz16}
Di, S., Cappello, F.: Fast error-bounded lossy {HPC} data compression with
  {SZ}. In: 2016 {IEEE} International Parallel and Distributed Processing
  Symposium, {IPDPS} 2016, Chicago, IL, USA, May 23-27, 2016. pp. 730--739
  (2016)

\bibitem{glecler}
Gleckler, P.J., Durack, P.J., Stouffer, R.J., Johnson, G.C., Forest, C.E.:
  Industrial-era global ocean heat uptake doubles in recent decades. Nature
  Climate Change  (2016)

\bibitem{habib2016hacc}
Habib, S., Pope, A., Finkel, H., Frontiere, N., Heitmann, K., Daniel, D.,
  Fasel, P., Morozov, V., Zagaris, G., Peterka, T., et~al.: Hacc: Simulating
  sky surveys on state-of-the-art supercomputing architectures. New Astronomy
  42,  49--65 (2016)

\bibitem{huffman}
Huffman, D.A., et~al.: A method for the construction of minimum-redundancy
  codes. Proceedings of the IRE  40(9),  1098--1101 (1952)

\bibitem{Kumar2013}
Kumar, A., Zhu, X., Tu, Y.C., Pandit, S.: Compression in molecular simulation
  datasets. In: International Conference on Intelligent Science and Big Data
  Engineering. pp. 22--29. Springer (2013)

\bibitem{isabela}
Lakshminarasimhan, S., Shah, N., Ethier, S., Ku, S., Chang, C., Klasky, S.,
  Latham, R., Ross, R.B., Samatova, N.F.: {ISABELA} for effective in situ
  compression of scientific data. Concurrency and Computation: Practice and
  Experience  25(4),  524--540 (2013)

\bibitem{zfp}
Lindstrom, P.: Fixed-rate compressed floating-point arrays. IEEE transactions
  on visualization and computer graphics  20(12),  2674--2683 (2014)

\bibitem{fpzip}
Lindstrom, P., Isenburg, M.: Fast and efficient compression of floating-point
  data. TVCG  12(5),  1245--1250 (2006)

\bibitem{ed2006}
Meyer, T., Ferrer-Costa, C., P\'{e}rez, A., Rueda, M., Bidon-Chanal, A., Luque,
  F.J., Laughton, C., Orozco, M.: Essential dynamics:  a tool for efficient
  trajectory compression and management. Journal of Chemical Theory and
  Computation  2(2),  251--258 (Mar 2006)

\bibitem{cpc2000}
Omeltchenko, A., Campbell, T.J., Kalia, R.K., Liu, X., Nakano, A., Vashishta,
  P.: Scalable i/o of large-scale molecular dynamics simulations: A
  data-compression algorithm. Computer Physics Communications  131(1),  78 --
  85 (2000)

\bibitem{ratana}
Ratanaworabhan, P., Ke, J., Burtscher, M.: Fast lossless compression of
  scientific floating-point data. In: Data Compression Conference, 2006. DCC
  2006. Proceedings. pp. 133--142. IEEE (2006)

\bibitem{ssem}
Sasaki, N., Sato, K., Endo, T., Matsuoka, S.: Exploration of lossy compression
  for application-level checkpoint/restart. In: Parallel and Distributed
  Processing Symposium (IPDPS), 2015 IEEE International. pp. 914--922. IEEE
  (2015)

\bibitem{sz17}
Tao, D., Di, S., Chen, Z., Cappello, F.: Significantly improving lossy
  compression for scientific data sets based on multidimensional prediction and
  error-controlled quantization. In: 2017 IEEE International Parallel and
  Distributed Processing Symposium, {IPDPS} 2017, Orlando, Florida, USA, May
  29-June 2, 2017. pp. 1129--1139 (2017)

\bibitem{Yang-sc1999}
Yang, D.Y., Grama, A., Sarin, V.: Bounded-error compression of particle data
  from hierarchical approximate methods. In: Proceedings of the 1999 ACM/IEEE
  Conference on Supercomputing. SC '99, ACM, New York, NY, USA (1999)

\end{thebibliography}

\end{document}